\documentclass[conference,compsoc]{IEEEtran}
\pdfoutput=1

\usepackage[nocompress]{cite}
\usepackage{graphicx}
\usepackage{booktabs}
\usepackage{multicol}
\usepackage{balance}
\usepackage{enumitem}
\usepackage[super]{nth}
\usepackage[flushleft]{threeparttable}

\usepackage{url}

\usepackage{xspace}
\newcommand{\mg}{MessageGuard\xspace}

\newcommand*\rot{\rotatebox{90}}

\usepackage{wasysym}
\newcommand{\full}{\CIRCLE}
\newcommand{\half}{\LEFTcircle}
\newcommand{\none}{\Circle}

\usepackage{color}
\definecolor{lightgray}{rgb}{.9,.9,.9}
\definecolor{darkgray}{rgb}{.4,.4,.4}
\definecolor{purple}{rgb}{0.65, 0.12, 0.82}

\usepackage{listings}
\usepackage{textcomp}
\lstdefinelanguage{JavaScript}{
  keywords={break, case, catch, continue, debugger, default, delete, do, else, false, finally, for, function, if, in, instanceof, new, null, return, switch, this, throw, true, try, typeof, var, void, while, with},
  morecomment=[l]{//},
  morecomment=[s]{/*}{*/},
  morestring=[b]',
  morestring=[b]",
  ndkeywords={class, export, boolean, throw, implements, import, this},
  keywordstyle=\color{blue}\bfseries,
  ndkeywordstyle=\color{darkgray}\bfseries,
  identifierstyle=\color{black},
  commentstyle=\color{purple}\ttfamily,
  stringstyle=\color{red}\ttfamily,
  sensitive=true
}

\emergencystretch=\maxdimen
\begin{document}

\title{MessageGuard: A Browser-based Platform for\\Usable, Content-Based Encryption Research}

\author{
\IEEEauthorblockN{Scott Ruoti\raisebox{4pt}{$\ast$}\raisebox{4pt}{$\dagger$}, Jeff Andersen\raisebox{4pt}{$\ast$}, Tyler Monson\raisebox{4pt}{$\ast$}, Daniel Zappala\raisebox{4pt}{$\ast$}, Kent Seamons\raisebox{4pt}{$\ast$}}
\IEEEauthorblockA{Brigham Young University\raisebox{4pt}{$\ast$}, Sandia National Laboratories\raisebox{4pt}{$\dagger$}\\
\{ruoti, andersen, monson\}@isrl.byu.edu, \{zappala, seamons\}@cs.byu.edu}
}

\maketitle

\begin{abstract}
%  Users share large amount of private data on the web, but very little
%  of it is protected with end-to-end encryption. One of the major
%  obstacles to wider adoption of encryption is the poor usability of
%  existing tools. Usable security research has tried to remedy this
%  problem, but to date most research has been ad hoc, with
%  improvements made to individual systems, and very little comparative
%  research among systems. 
  
This paper describes MessageGuard, a browser-based platform for research into usable content-based encryption.  
MessageGuard is designed to enable collaboration between security and usability researchers on long-standing research questions in this area.
It significantly simplifies the effort required to work in this space and provides a place for research results to be shared, replicated, and compared with minimal confounding factors.
MessageGuard provides ubiquitous encryption and secure cryptographic operations, enabling research on any existing web application, with realistic usability studies on a secure platform.
We validate MessageGuard's compatibility and performance, and we illustrate its utility with case   studies for Gmail and Facebook Chat. 
\end{abstract}

\section{Introduction}

Users share private information on the web through a variety of applications, such as email, instant messaging, social media, and document sharing.
HTTPS protects this information during transmission, but does not protect users' data while at rest.
Additionally, middleboxes can weaken HTTPS connections by failing to properly implement TLS or adequately validate certificate chains~\cite{liang2014https}.
Even if a website correctly employs HTTPS and encrypts user data while at rest, the user's data is still vulnerable to honest-but-curious data mining~\cite{ruoti2013confused}, third-party library misbehavior~\cite{stefan2014protecting}, website hacking, protocol attacks~\cite{foster2015security,durumeric2015neither,holz2016tls}, and government subpoena.

This state of affairs motivates the need for {\em content-based} encryption of user data.
%, rather than relying on the connection-based encryption afforded by protocols like HTTPS.
In content-based encryption, users' sensitive data is encrypted at their own computer and only decrypted once it reaches the intended recipient, remaining opaque to the websites that store or transmit this encrypted data.
The best known examples of content-based encryption are secure email (e.g., PGP, S/MIME) and secure instant messaging (e.g., OTR).
In addition to protecting communication, content-based encryption can protect any data users store or distribute online; for example, files stored in the cloud (e.g., DropBox, Google Drive) or tasks in a web-based to-do list.

Unfortunately, while there is a large body of work on messaging protocols and key management schemes for content-based encryption~\cite{unger2015sok}, too little of it has been subjected to formal usability evaluation.
This is problematic as experience has shown that many proposals with strong theoretical foundations are either unfeasible in real-world situations~\cite{garrison2016practicality} or have problems that are only apparent when studied empirically~\cite{whitten1999why,dhamija2005battle,ruoti2015authentication}.
As a result, there are many open HCI and security questions regarding content-based encryption~\cite{garfinkel2014usable}.

\subsection{MessageGuard}

To address these concerns, we have developed \mg, a platform for usable security research focused on content-based encryption.
Currently, participating in research of this area has been a costly affair, requiring researchers to either build a system from scratch~\cite{fahl2012confidentiality,he2014shadowcrypt} or to make substantial modifications to one of the existing open source tools~\cite{atwater2015leading}.\footnote{Substantial modifications are needed because existing open source content-based encryption tools are unusable~\cite{ruoti2015why}.}
\mg is designed to greatly simplify the effort required to develop usable content-based encryption
Ultimately, we hope that \mg will enable collaboration between applied cryptography and usability researchers, allowing them to solve the long-standing open questions in this field, such as the creation of usable and secure key management.
% to finally solve the key management problems that have plagued security software for decades.

The benefits of \mg include:

\begin{enumerate}

\item {\em Accelerates the creation of content-based encryption prototypes.}
\mg provides a fully functional content-based encryption system, including user interfaces, messaging protocols, and key management schemes.
The modular design of \mg allows researchers to easily modify only the portions of the system they wish to experiment with, while the remaining portions continue operating as intended.
This simplifies development and allows researchers to focus on their areas of expertise --- either usability or security.

\item {\em Provides a platform for sharing research results.}
Researchers who create prototypes using \mg can share their specialized interfaces, protocols, or key management schemes as one or more patches,\footnote{A diff-based patch of \mg's code base.}, allowing other researchers to leverage and replicate their work.
Additionally, successful research that would benefit all \mg prototypes can be merged into \mg's code base, allowing the community to benefit from these advances and reducing fragmentation of efforts.%\footnote{This is similar to the benefit that LLVM has provided to the compiler research community.}

\item {\em Simplifies the comparison of competing designs.}
\mg can be used to rapidly develop prototypes for use in A/B testing~\cite{ruoti2016private}.
Two prototypes built using \mg will only differ in the areas that have been modified by researchers.
This helps limit the confounding factors that have proven problematic in past comparisons of content-based encryption systems~\cite{ruoti2013confused,atwater2015leading,ruoti2016we}.
The usability studies reported in this paper provide a baseline for comparison against future work built on \mg.

\item {\em Retrofits existing web applications with content-based encryption.}
  Because \mg works with all websites, in all browsers, and on both desktop and mobile platforms, it enables
  researchers to design usable content-based encryption for a wide variety of applications.
  Researchers do not need to cooperate with application developers or service providers, allowing them
  to easily work on systems that have large, installed user bases.  This removes a confounding factor
  when conducting user studies, since users will be familiar with the application they are using. It
  also enables long-term usability studies, with users interact with encryption as part of their
  daily habits.

\item {\em Provides secure cryptographic operations for all applications.}
  \mg uses security overlays~\cite{ruoti2013confused} to isolate a user's sensitive content from web applications, ensuring that only an encrypted copy of the user's data is available to those web applications.
Thus \mg enables HCI researchers to easily test their ideas with applications that are actually secure, rather
than relying on mock-systems, which could run the risk of invalidating their results.

\end{enumerate}

In this paper we describe the \mg platform, including our threat model, system goals, and implementation details.
Our contributions include the following items. (1) A description of the \mg platform and a guide for how researchers can use it to develop usable, content-based encryption for existing web applications.
(2) Development of the first content-based encryption system that is designed to work with all websites, in all browsers, and on both desktop and mobile platforms.
(3) A validation of \mg's compatibility and performance, showing that it supports all of the Alexa Top 50 (e.g., Facebook, Instagram, Twitter) websites with little impact on page load times.
(4) Two case studies that illustrate how \mg meets its design objectives, including six usability studies with 203 participants demonstrating that users find the prototypes we built using \mg to be highly usable.

\subsection{Areas of Potential Research}

\mg enables researchers to examine diverse applications of content-based encryption in web applications.
For example, \mg can be used to improve existing forms of content-based encryption such as secure email or secure chat.
Alternatively, researchers could use \mg to evaluate the feasibility of adding content-based encryption to web applications in novel contexts,
such as signing Tweets for highly-targeted Twitter accounts or securing cloud storage.

Researchers can use \mg to implement new interfaces, protocols, algorithms, and key management schemes while leveraging \mg's existing usable interfaces and security features.
Prototypes built with these new research features can be evaluated empirically, measuring the effect that they have on the user experience.
%Additionally, these prototypes allow real-world measurements of a new feature's performance.

\newcounter{ideas}\stepcounter{ideas}

Potential areas of exploration and collaboration for security and HCI researchers include, but are not limited to:
(\alph{ideas}\stepcounter{ideas}) designing content-based encryption interfaces that are resilient to spoofing;
(\alph{ideas}\stepcounter{ideas}) exploring messaging protocols and key management schemes for content-based encryption; for example, certificate revocation, certificate transparency, key ratcheting, puncturable encryption;
(\alph{ideas}\stepcounter{ideas}) developing instructive interfaces that help users build correct mental models of content-based encryption;
(\alph{ideas}\stepcounter{ideas}) investigating key escalation, which starts users on an easy-to-use, but less secure form of key management (e.g., passwords) and then migrates them to a more secure key management scheme (e.g., PGP) as they gain expertise;
(\alph{ideas}\stepcounter{ideas}) creating interfaces that help users avoid mistakenly sending sensitive data in the clear;
(\alph{ideas}\stepcounter{ideas}) supporting easy-to-use key discovery for traditional public key cryptography (e.g., PGP);
(\alph{ideas}\stepcounter{ideas}) notifying users of potential insecurities regarding their encrypted content; and
(\alph{ideas}\stepcounter{ideas}) assisting users in migrating their encryption keys between devices they own.

\section{Related Work}

Unger et al.~\cite{unger2015sok} provide a comprehensive overview of secure messaging protocols, which use content-based encryption,
and discuss a variety of key management schemes.
Unger et al.'s work demonstrates that there is strong interest in content-based encryption within the security community.
Still, examining the systems highlighted by Unger et al.'s survey makes it clear that few proposals have ever undergone user studies, making clear the need for a platform to easily prototype and test these proposals.
%making it unclear if these proposals are actually helpful to end users.
%This is especially troubling, as previous work has shown a disconnect between what security experts consider to be a good system and what users actually consider to be usable~\cite{whitten1999why,dhamija2005battle,ruoti2015authentication}.
%This highlights the need for engagement by the HCI community in end-to-end encryption research.

\subsection{Security Overlays}

There have been several systems that have used security overlays~\cite{van2009encrypted,ruoti2013confused} to enhance existing web applications with content-based encryption:
Fahl et al.~\cite{fahl2012confidentiality} describe Confidentiality-as-a-Service (CaaS), a system designed to make it easy for users to encrypt their sensitive data stored in the cloud. 
%CaaS is designed to encrypt data so the cloud provider is unable to access the plaintext. 
%To increase usability, key management is entirely transparent to the end users.
CaaS uses Greasemonkey, a Mozilla Firefox extension, to add encryption capabilities to existing web pages.
%The Greasemonkey sandbox prevents the web service provider from accessing the plaintext. 
The paper describes proof-of-concept implementations of CaaS for Dropbox, Facebook, and email.
%Usability studies demonstrate that the system is highly usable.

%Ruoti et al.~\cite{ruoti2013confused} present Pwm, a system designed for easy encryption of webmail.
%It utilizes \texttt{iframes} to secure sensitive content, similar to \mg.
%Like CaaS, it also supports automatic key management and has been shown to have high usability.
%Unlike other systems, Pwm focuses on helping first time recipients of an encrypted email understand how to decrypt the message and begin using Pwm.

He et al.~\cite{he2014shadowcrypt} proposed ShadowCrypt, a Google Chrome extension that sits between the user and their web services and allows the users to create and consume encrypted content without revealing the content to the web service.
ShadowCrypt displays encrypted contents using the Shadow DOM, an upcoming \texttt{HTML5} standard.\footnote{Only blink-based browsers support the Shadow DOM.}
%Currently the Shadow DOM is only available in Blink-based browsers (i.e., Chrome and Opera). Blinks implementation of the Shadow DOM is still incomplete.}
%, where it is still only partially implemented.}
%The Shadow DOM allows applications to define a ``shadow'' document fragment (i.e., \texttt{ShadowRoot}) that will be rendered in place of a specified portion of the document's actual \texttt{DOM} (i.e., \texttt{host}).
%Elements defined in the \texttt{ShadowRoot} are invisible to most \texttt{DOM} methods and can only be accessed through the \texttt{ShadowRoot}.
%For example \texttt{document.getElementById} will not find elements defined in the \texttt{ShadowRoot}, instead you need call \texttt{getElementById} on the \texttt{ShadowRoot} object.
Unfortunately, as discussed in Appendix \S1, several flaws compromise the security of this approach.
%the use of the Shadow DOM to secure users' content is ill-advised.

Lau et al.~\cite{lau2014mimesis} present Mimesis Aegis (M-Aegis), a privacy-preserving system for mobile platforms.
%Three of M-Aegis goals are very similar to our work:
%First, it protects sensitive data from untrusted entities, including applications running on the mobile platform.
%Second, it preserves the user experience with existing apps rather than require the user to move to a separate secure app.
%Third, it attempts to create a model that will generalize to any number of applications.
M-Aegis places a transparent window on top of the application GUI in order to intercept and encrypt data before it reaches the native application.
M-Aegis uses a novel design that utilizes features of the accessibility layer of the operating system in order to overlay the interface of any application.
To our knowledge, M-Aegis is the first system that attempts to provide ubiquitous and integrated encryption outside of the browser.
A prototype implementation of M-Aegis overlaying Gmail was part of a study with 15 participants that showed most of the participants did not report any noticeable difference between the original app and the app with M-Aegis enabled.

In comparison to these systems, \mg is the first system that is capable of supporting content-based encryption using security overlays across all desktop and mobile platforms.
A more in-depth evaluation of these systems, their strengths and weaknesses can be found in Appendix~\ref{appx:strategies}.

\subsection{Other approaches}

In contrast to using security overlays, there are systems that attempt to provide content-based encryption through a combination of modifying the browser and modifying existing web applications.

In this first category, COWL modifies the JavaScript runtime to provide confinement between different scripts, enforcing mandatory access
control~\cite{stefan2014protecting}. For example, this allows untrusted JavaScript, such as a third-party library, to compute over sensitive
data but not to transmit that data to an untrusted server. This provides powerful capabilities to the browser,
enabling content-based encryption to be widely supported, but requires modifications to JavaScript and cooperation
from the application provider. Hails provides a similar confinement system, written in Haskell~\cite{giffin2012hails}.
Approaches that require a new runtime are still largely theoretical and may struggle to cope with complex web applications.

In this second category, Content Cloaking provides a browser extension recognizes and intercepts AJAX requests that
the web application makes to the content provider.  The extension then encrypts data as it leaves the browser
and is sent to a content-provider, and then decrypts data when it arrives
from the provider and before it is displayed by the browser. This must be customized for each web application, and was only
demonstrated for Google Docs. Similarly, Beeswax requires tight cooperation
between application developers and the security platform
\cite{legare2016beeswax}. Developers must indicate which DOM elements
should be kept private and which users can share the contents of those
elements, and then the platform provides cryptographic operations and
key management.

In comparison to these systems, \mg does not require cooperation from web applications.
This is an important distinction for two reasons: (1) modifying individual applications does not scale with the explosive growth of web applications, and (2) requiring cooperation from web developers severely limits the ability of researchers to explore content-based encryption in the applications that most interest them.
For these reasons, in this paper we focus on adding content-based encryption using secure overlays.

\section{Ubiquitous, Content-Based Encryption}
In this section, we give the threat model that motivates our work.
Next, we describe how security overlays can be used to enhance existing web applications with content-based encryption.
Finally, we discuss our goals for \mg, that are necessary to support research of content-based encryption in a usable, secure, and extensible manner.
% the web with content-based encryption in a way that could be adopted by the masses.
%Next, we detail where content-based encryption needs to be implemented in order to be compatible with existing browsers.
%Finally, we discuss limitations inherent to retrofitting the web with content-based encryption.

\subsection{Threat Model}
In content-based encryption, sensitive content is only accessible to the author of that data and the intended recipient.
In contrast to transport-level encryption (e.g., TLS), which only protects data during transit, content-based encryption protects data both during transit and while it is at rest.
In our threat model, we consider web applications, middleboxes (e.g. CDNs), and the content they serve to be within the control of the adversary.
The adversary wins if she is able to use these resources to access the user's encrypted data.
%any sensitive data that is accessible to the adversary, either directly as part of the DOM or through JavaScript, is considered compromised.
While it is true that most websites are not malicious,\footnote{Many websites are best described as honest-but-curious.} in order to support ubiquitous, content-based encryption, it is necessary to protect against cases where websites are actively trying to steal user content.
%retrofit the web with content-based encryption it is necessary to also address those websites that are truly malicious.
%must also handle websites, middleboxes, and network infrastructure which truly are malicious.
Users' computers, operating systems, software, and content-based encryption software\footnote{This includes the software's website and any web services the software relies upon (e.g., a key server).} are all considered part of the trusted computing base in our threat model.

Our threat model is concerned with ensuring the confidentiality, integrity, and authenticity of encrypted data, but does allow for the leakage of meta-data necessary for the encrypted data to be transmitted and/or stored by the underlying web application.
For example, in order to transmit an encrypted email message, the webmail system must have access to the unencrypted email addresses of the message's recipient.
Additionally, the webmail provider will be able to inspect the encrypted package and gain learn basic information about the encrypted package (e.g., approximate length of message, number of recipients).\footnote{This type of leakage also occurs in HTTPS.}

While our threat model is necessarily strict to support the wide range of web applications that researchers may wish to investigate, we note that research prototypes built using the \mg platform are free to adopt a weaker threat model that may be more appropriate for that particular research.

\subsection{Security Overlays}
\begin{figure}[t]
\begin{center}
\includegraphics[width=1.0\columnwidth]{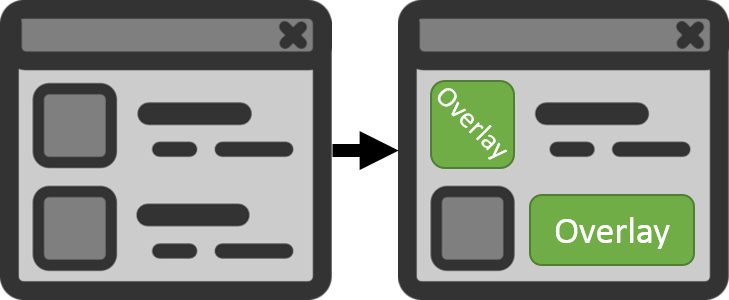}
\end{center}

\vspace{.5\baselineskip}Portions of the web application (shown on the left) have been overlayed with secure interfaces (shown on the right).
\vspace{.5\baselineskip}

\caption{Overlaying a web application}
\label{fig:overlays}
\end{figure}

To encrypt user data before it reaches web applications, we leverage a technique known as \textbf{overlaying}~\cite{van2009encrypted,ruoti2013confused}.
In this technique, \mg replaces portions of the web application's interface with secure interfaces known as \textbf{overlays} (see Figure~\ref{fig:overlays}).\footnote{The overlayed elements are not actually removed, but visually occluded by the secure overlays.}
While a security overlay appears to be a part of the web application, the security overlay itself is inaccessible to the web application.

There are several approaches for implementing overlays: \texttt{iFrames}~\cite{van2009encrypted,ruoti2013confused}, the \texttt{ShadowDOM}~\cite{he2014shadowcrypt}, Greasemonkey~\cite{fahl2012confidentiality}, and the operating system's accessibility framework~\cite{lau2014mimesis}.
An in-depth analysis of each of these approaches can be found in Appendix \S1.
Based on our analysis of each of these approaches, \texttt{iFrames} are the implementation strategy best suited to work across all operating systems and browsers (including mobile).
Additionally, \texttt{iFrame}-based security overlays have security and usability that are greater than or equal to that of other approaches.
As such, we designed \mg using security overlays based on \texttt{iFrames}.

%Is this necessary or appropriate?
Relying on \texttt{iFrames} largely restrict \mg to supporting only web applications deployed in the browser.
Still the browser is an ideal location for studying content-based encryption:
(1) There are a large number of high usage browser-based web applications (e.g., webmail, Google Docs).
(2) Traditional desktop and mobile application development increasingly mimics web development, allowing lessons learned in browser-based researcher to also apply to these other platforms.
(3) There is already a substantial amount of research into adding content-based encryption to web applications, both academic
(e.g.,~\cite{fahl2012confidentiality,ruoti2013confused,he2014shadowcrypt,atwater2015leading}) and professional (e.g., Virtru, Mailvelope, Encipher.it).

\subsection{Platform Goals}
We examined the existing work on content-based encryption (e.g.,~\cite{whitten1999why,garfinkel2005johnny,sheng2006why,unger2015sok}) in order to establish a set of design goals for \mg.
These goals are centered around enabling to researcher to conduct research into usable, content-based encryption.

\subsubsection{Secure}
\mg should secure users' sensitive content from web applications and network adversaries.

{\em \mg should protect data in its overlays from being accessed by the web application.}
Sensitive data that is being created or consumed using \mg should be inaccessible to the web application which \mg has secured.
A corollary to this rule is that no entities that observe the transmission of data encrypted by \mg should be able to decipher that data unless they are the intended recipients.

{\em \mg's interfaces should be clearly distinguishable from the web application's interfaces.}
In addition to protecting content-based messages from websites, it is important that systems clearly delineate which interfaces belong to the website and which belong to the content-based encryption software.
This helps users to feel assured that their data is being protected and assists them in avoiding mistakes~\cite{ruoti2013confused,ruoti2016private}.
Additionally, visual indicators should be included that can help protect against an adversary that attempts to social engineer a user into believing they are entering text into a secure interface when in reality they are entering text directly into the adversary's interface~\cite{dhamija2005battle,bravo2012operating}.

\subsubsection{Usable}
\mg should provide a usable base for future research efforts.

{\em \mg should be approachable to novice users.}
Easy-to-use systems are more likely to be adopted by the public at large \cite{unger2015sok}.
Furthermore, complicated systems foster user errors, decreasing system security \cite{whitten1999why,ruoti2013confused}.
While some systems need to expose users to complex security choices, basic functionality (e.g., sending or receiving an encrypted email) should be approachable to new users.
At a minimum this includes building intuitive interfaces, providing integrated, context-sensitive tutorials, and helping first time recipients of encrypted messages understand what they need to do in order to decrypt their message.

{\em \mg should integrate with existing web applications.}
Users enjoy the web services and applications they are currently using and are disinclined to adopt a new system solely because it offers greater security.
Instead, users prefer that content-based encryption be integrated into their existing applications~\cite{ruoti2013confused,atwater2015leading}.
Equally important, content-based encryption should have a minimal effect on the application's user experience; if encryption gets in the way of users completing tasks it is more likely that they will turn off content-based encryption~\cite{herley2009so}.

{\em \mg's interfaces should be usable at any size.}
Current web interfaces allowing users to consume or create content come in a wide variety of sizes (i.e., height and width).
When \mg integrates with these web services, it is important that \mg's interfaces work at these same sizes.
To support the widest range of sizes, \mg's interfaces should react to the space available, providing as much functionality as is possible at that display size.

%{\em \mg Users should control when their data is encrypted.}
%Research has shown that most users are not interested in encrypting all of their online data~\cite{gaw2006secrecy,ruoti2013confused}.
%As such, users should be able to select which information will be encrypted and which will be transmitted in the clear.

\subsubsection{Ubiquitous}
\mg should support most websites and platforms.

{\em \mg should work with most websites}
\mg should make it easy for researchers to explore adding end-to-end encryption into whichever web applications they are interested in.
While it may be impossible to fully support all web applications (e.g., Flash applications or applications drawn using an HTML canvas), most standard web applications should work out-of-the-box.
For those applications which don't work out-of-the-box, \mg should allow researchers to create customized prototypes that handle these edge cases.

{\em \mg should function in all major desktop and mobile browsers.}
Prototypes built with \mg should function both on desktop and mobile browsers, allowing researchers to experiment with both of these form factors.
Furthermore, \mg should work on all major browsers, allowing users to work with the web browser they are most familiar with, obviating the need to restrict study recruitment to users of a specific browser.

\subsubsection{Extensible}
\mg should be easily extensible and contribute to the rapid development of content-based encryption prototypes.

{\em \mg should be modular.}
\mg's functionality should be split into a variety of modules, with each module taking care of a specific function.
Researchers should also be free to only change the modules that relate to their research, and have the system continue to function as expected.
Similarly, \mg's modules should be extensible, allowing researchers to create new custom modules with a minimal amount of effort.

{\em \mg should provide reference functionality.}
As a base for other researchers' work, \mg should include a reference implementation of the various modules that adds content-based encryption to a wide range of web applications.
This reference implementation should be able to be easily modified and extended to allow researchers to rapidly implement their own ideas.
%Furthermore, it should be well commented, allowing researchers to understand how to use \mg.

\subsubsection{Reliable}
The usability and security of \mg should be reliable, protecting researchers from unintentionally compromising \mg's security or usability.

{\em Reducing the security of \mg should require deliberate intent.}
HCI researchers should feel comfortable customizing \mg's interface without needing to worry that they are compromising security.
To facilitate this, \mg should separate UI and security functionality into separate components.
As long as researchers limit themselves to changing only UI components, there should be no effect on security.

{\em Modifying the cryptographic primitives should have minimal effect of \mg's usability.}
As above, \mg should separate its UI and security functionality into separate components.
This will allow security researchers to modify the cryptographic primitives without worrying about how they will affect \mg's usability.
The one caveat to this is if a new key management scheme requires a user interface that \mg does not already make available.
In this case, researchers will need to provide this key management scheme's interface, which could affect usability, but other interfaces should remain unaffected.

\section{\mg}
\begin{figure}
\begin{center}
\includegraphics[width=\columnwidth]{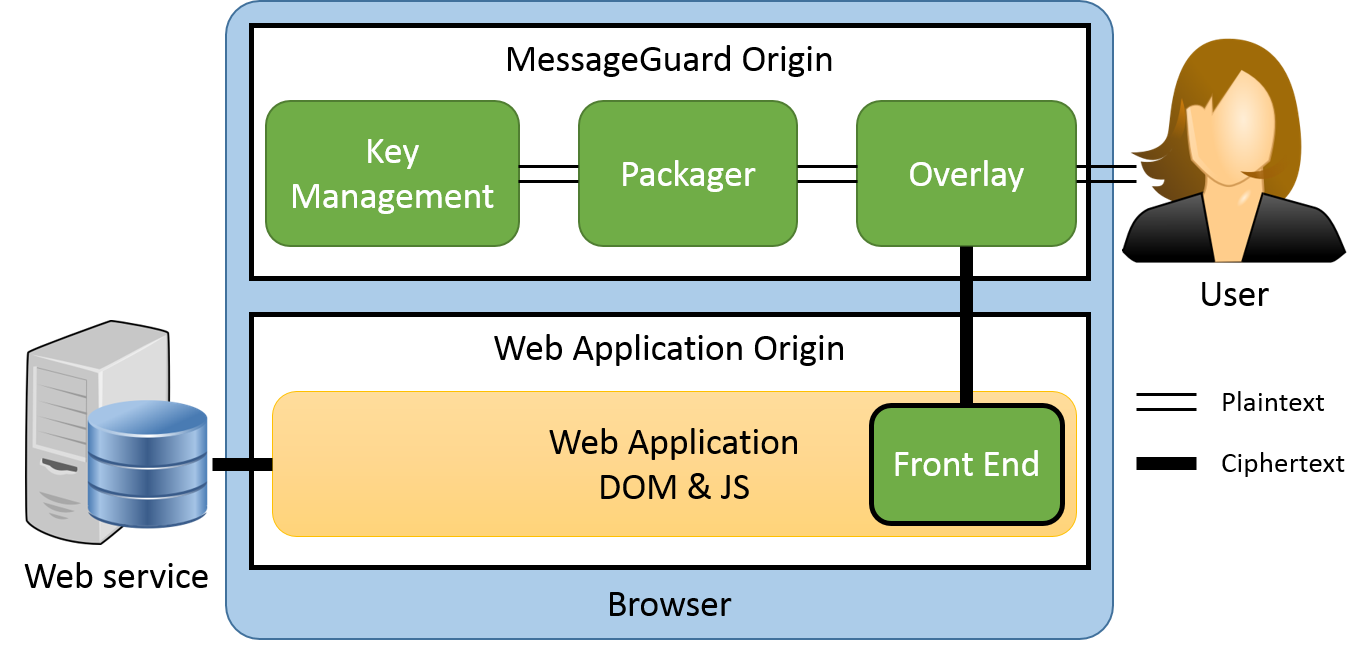}
\end{center}

\vspace{.5\baselineskip}
{\small User's sensitive data is only accessible within the \mg origin.}
\vspace{.5\baselineskip}

\caption{Overview of \mg architecture.}
\label{fig:architecture}
\end{figure}

Based on our design goals and using \texttt{iFrame}-based security overlays we created \mg.
%\mg is the first system that retrofits the entire web with content-based encryption and works with all websites, in all browsers, on all platforms, both desktop and mobile. 
%It is also the only system that attempts to address all of the system goals we identified for content-based encryption (Section~\ref{sec:goals}).
Figure~\ref{fig:architecture} shows an overview of \mg's architecture.
%
%Since \mg is a platform, each component (i.e., front end, overlay, packager, and key management) has been designed to by customizable by researchers on a global, on a per-application-type (e.g., email, chat), or a per-application (e.g., Twitter, Flicker) basis.

All source code related to \mg can be found at \url{https://bitbucket.org/isrlemail/messageguard}.
An example secure email prototype built using \mg can be found at \url{https://messageguard.io}.\footnote{
This prototype secures email sent and received through Gmail and uses PGP-based key management. Similar prototypes that secure email using passwords or identity-based encryption are also available upon request.}

In the remainder of this section, we first describe \mg's workflow.
We then detail the design of \mg's core components, and describe \mg's default functionality.\footnote{Per-application customization is discussed in \S\ref{sec:platform}.}
We then discuss several web service (i.e., key server, encrypted file server) we have created to further aid researchers in building solutions using \mg. 
Finally, we give a brief overview of \mg's implementation.

\subsection{Workflow}
\mg's workflow is as follows:

\begin{enumerate}

\item \mg injects the \textbf{front end} component into the website's front end. \mg's front end component scans for encrypted payloads and data entry interfaces. When found, it replaces these items with an \textbf{overlay}. After this initial scan, changes to the page are tracked and only elements that have been modified are scanned.

\item
Read overlays are used to display sensitive information to the user, and a compose overlay allows users to encrypt sensitive information before sending it to the website.
Each overlay is displayed within an \texttt{iframe} and uses the browser's same-origin policy to protect its contents from the website.
Overlays use the \textbf{packager} component to handle the actual encryption/decryption and packaging of data.

\item 
In conjunction with the \textbf{key management} component, the packager encrypts and decrypts data.
The packager also encodes encrypted data, making it suitable for transmission to and from the website's front end.
\end{enumerate}

\subsection{Front End}

\begin{figure}[t]
\centering
\includegraphics[width=\columnwidth]{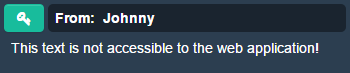}
\caption{Generic Read overlay.}
\label{fig:read-overlay}
\end{figure}

\begin{figure}[t]
\centering
\includegraphics[width=\columnwidth]{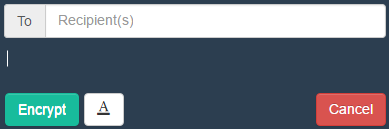}
\caption4{Generic Compose overlay.}
\label{fig:compose-overlay}
\end{figure}

\mg's front end component is injected into every website.
The front end has three responsibilities:

\begin{enumerate}

\item \textbf{Identify encrypted payloads.}
The front end identifies encrypted payloads, overlays these payloads with a read overlay, and sends the payloads to the overlay so that they can be decrypted and displayed to the user (Figure~\ref{fig:read-overlay}).

\item \textbf{Identify data entry interfaces.}
The front end identifies interfaces where the user may wish to enter encrypted data (e.g., \texttt{[contentEditable]} and \texttt{textarea} elements).
Each of these interfaces are then modified to display a button that users can click to replace the interface with a compose overlay (Figure~\ref{fig:compose-overlay}).%\footnote{Usability studies have shown that most users prefer not to encrypt content by default \cite{ruoti2013confused,ruoti2015johnny}.}
The front end also creates an overlay manager for each overlay that handles passing data between the the compose overlay and the website (e.g., sending the overlay the encrypted payload, saving drafts).

\item \textbf{Displaying context-sensitive tutorials.}
The front end displays tutorials that instruct new users how to use \mg. These are all context-sensitive, appearing as the user performs a given task for the first time.

\end{enumerate}

The front end is the only \mg component that runs outside of \mg's protected origin.
%can be corrupted by the website it is injected into.
For this reason, \mg is designed to treat the front end as untrusted.
Since the front end component does not exist as part of \mg's protected origin, it cannot directly access the packager, key management, or overlay components.
Instead, it is limited to communicating with the overlay component using the web messaging API~\cite{spec:webMessaging}.
Additionally, the overlay always encrypts user data before transmitting it to the front end component and sanitizes any data it receives from the front end.

\subsubsection{Default Functionality}
\mg's default front end modifies all web applications to allow for content-based encryption.
The default front end will create a read overlay for encrypted payloads found anywhere on the page.
It will also allow encryption in all larger textual entry interface elements (i.e., \texttt{[contentEditable]} and \texttt{textarea} elements).
The front end also contains generic tutorials, which are shown when users first encounter new functionality in \mg.

\subsection{Overlays}
\mg's overlays are designed to mimic the placement and dimensions of the content they overlay and to be visually appealing and intuitive.
Overlays have a distinctive, dark color scheme that stands out from most websites, allowing users to easily identify secure overlays from insecure website interfaces.
%There is also a user-selected watermark displayed in each overlay that further differentiates it from interfaces created by the website,\footnote{For readability, the watermark is not shown in the figures.} though research into potentially better indicators is ongoing.
Finally, overlays sanitize the plaintext contents of encrypted messages in order to prevent malicious messages from compromising the overlay.

\subsubsection{Default Functionality}

\begin{figure}[t]
\centering
\includegraphics[width=\columnwidth]{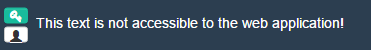}
\caption{Generic read overlay in constrained space.}
\label{fig:tiny-read}
\end{figure}

\begin{figure}[t]
\centering
\includegraphics[width=\columnwidth]{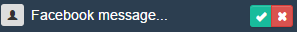}
\caption{Generic compose overlay in constrained space.}
\label{fig:tiny-compose}
\end{figure}

\mg includes two standard overlays: a read overlay (Figure~\ref{fig:read-overlay}) and a compose overlay (Figure~\ref{fig:compose-overlay}).\footnote{Support for a file upload overlay is also nearing completion.}
Read overlays are responsible for decrypting and displaying sensitive information to users.
Compose overlays allow users to compose rich text messages and encrypt these messages before sending them to the website.
Both overlays are reactive and modify their layout based on available space.
For example, at small sizes, the compose overlay no longer shows tools for formatting text, but still allows their use through keyboard shortcuts (see Figure~\ref{fig:tiny-read} and Figure~\ref{fig:tiny-compose}).

\subsection{Packager}
The packager encrypts/decrypts user data and encodes the encrypted data to make it suitable for transmission through web applications.
The packager uses standard cryptographic primitives and techniques to encrypt/decrypt data (e.g., AES-GCM).
Ciphertext is packaged with all information necessary for recipients of the message to decrypt it.

\subsubsection{Default Functionality}
\mg's default packager is based on the Cryptographic Message Syntax (CMS)~\cite{housley1999cryptographic} used in S/MIME.
It differs from CMS in that all non-essential attributes (e.g., versioning information) are removed.
This was done to reduce the package size, which is necessary for \mg to support web applications that constrain the size of packages (e.g., chat applications).
%To further decrease the size of packages, message contents are zipped before encryption.
%Instead of using a verbose packaging format (e.g., XML, JSON), we use the ProtoBuf packaging format to reduce the size of the messages and speed up processing.\footnote{As part of preparing the data for transmission, we also base64 encode the binary data produced by the ProtoBuf library.}

\subsection{Key Management}
\label{sec:keyManagement}
\mg's key management component provides a UI (displayed in \mg's options page) that lists the keys currently available to the user and allows the users to create, register, and delete keys.
The key management component also manages storage for the various key management schemes.
It includes encrypted storage for sensitive information (e.g., private keys), that is protected by a master password set by the user when first running \mg.
Finally, we note that while all the prototypes we have built with \mg have only used a single key management scheme at a time, \mg supports prototypes that permit users to pick and choose which key management schemes they want to use for which messages.

\subsubsection{Default Functionality}
We have created three reference key management schemes for use in \mg:

\begin{figure}[t]
\centering
\includegraphics[width=1.0\columnwidth]{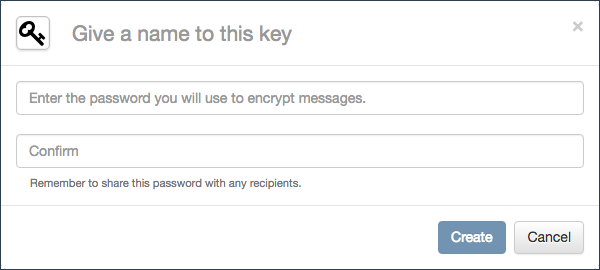}
\caption{Interface for creating keys using a shared password.}
\label{fig:key-ss}
\end{figure}

\begin{itemize}

\item \textbf{PGP.}
We have created a standard-compliant implementation of PGP.
This implementation generates 2048-bit keys for users, and publishes these keys for discovery on the \mg's key server web service.
The PGP private key pair is stored in the user's browser, but is encrypted using the key manager's master password.

\item \textbf{Identity-based encryption (IBE).}
We have implemented the Boneh-Boyen IBE scheme~\cite{boneh2004efficient} for use in \mg.
The key server web service is responsible for storing the system parameters and master secret.
Clients automatically generate public keys based on recipient identifiers.
The private key is retrieved from the key server and stored in the user's browser, but is encrypted using the key manager's master password.

\item \textbf{Shared Password.}
Finally, we have created a key management scheme where emails are encrypted using passwords shared between users.
These passwords are transformed into key material through the use of the PBKDF2 function.
This key management scheme does not use the key server, and users are responsible for sharing their passwords out of band.
The derived key material can either be stored in the user's browser and encrypted by the key manager's master password, or users can re-enter their shared passwords each time they want to encrypt or decrypt a message.\footnote{We have created one key management scheme for each behavior.}

\end{itemize}

Figure~\ref{fig:key-ss} is an example of the interfaces users see when adding a key to \mg.

\subsection{Web Services}
To help researchers in creating systems using \mg, we have created two web services that inter-operate with \mg: a key server and a file server.
The code for both servers is in \mg's code repository, and researchers are welcome to have their prototypes point at our key server and file server, or deploy their own.

\subsubsection{Key Server}
To facilitate key management schemes that require discovery of public keys (e.g., PGP, S/MIME) or require key escrow (e.g., IBE) we have created a key server.
The key server requires that users create an account with a username and a password.
After account creation, users prove their ownership of other accounts (e.g., email, Twitter, Facebook).
Once a user has established ownership of an account, ownership cannot be transferred to another owner unless first released by the original owner.

This key server exposes two sets of REST endpoints: public and private.
The public endpoints are used to retrieve public data (e.g., public keys, IBE system parameters).
The private endpoints are limited to users that have proved ownership of the appropriate account (e.g., setting a public key for a given email address, retrieving private IBE keys, etc.).

Both the PGP and IBE key management schemes we have included with \mg make use of the key server. 
For the PGP scheme, users upload their public keys to the key server (via private endpoint), and other users download those public keys as needed (via public endpoint).
For IBE, users can query the key server for the system parameters (via public endpoint), and also retrieve their private keys (via private endpoint).
We welcome other researchers to use our publicly available key server, but they can also deploy their own as needed with the code hosted in the repository.

\subsubsection{File Server}
We have created a file server that allows for capability-based storage of files.
Currently, the file server allows anyone to upload a file, after which they are given a capability (i.e., UID) that can be used to access that file.
While we currently do not require authentication to the key server, that is something that could easily be added.

In practice, we use the file server to enable attachments in encrypted email.
In our work using \mg to build a secure integrated email prototype, we were not able to use the underlying email application's native file attachment functionality, and instead relied upon the file server to store encrypted attachments.
This was done by encrypting the attachment with a random key, and storing the encrypted file sans key on the file server.
The encryption key, a cryptographic hash of the attachment, and the capability for the attachment are stored in the encrypted email.
As such, only individuals who can decrypt the email message are able to access and decrypt the attachment.

\subsection{Implementation}
We implemented \mg so that it would run on all major desktop browsers (i.e., Chrome, Firefox, Internet Explorer, Opera, Safari) and mobile browsers (i.e., Android, iOS, Windows Phone).
We employed only standard JavaScript functions that were confirmed to work on all major browsers.\footnote{\url{http://caniuse.com/}.}
We avoided injecting polyfills into the web application and polluting the \texttt{window} object in an effort to ensure that \mg did not break existing web applications.
\mg is implemented as both a browser extension and as a bookmarklet (i.e., user script).

\mg has a single codebase, with only a small portion of code that is used to address differences in various browsers (<1\%).
This codebase is implemented in JavaScript, Sass, and HTML.
The code is compiled into browser extensions and a bookmarklet using NodeJS and Gulp.
The code itself is split into a number of JavaScript modules which are bundled together at compile time using Browserify.
Browserify also allows \mg to leverage a large number of quality NodeJS modules (e.g., CryptoJS, jQuery, Bootstrap).
The build system also generates source maps to facilitate debugging, runs style checking on the code, and generates documentation using JSDoc.
Instructions for setting up and building \mg are available in \mg's repository.

Further details regarding the technologies used to implement \mg can be found in Appendix \S2.

\section{\mg as a Research Platform}
\label{sec:platform}

In this section, we describe the ways researchers can employ \mg as a platform for their own research.
In addition to the details described in this section, we invite researchers to download \mg's source code.
To help researchers quickly familiarize themselves with \mg's code base, we have included instructive comments throughout the code and have provided a reference implementation that supports most websites.\footnote{The code base also includes an implementation of secure email as an additional reference.} that researchers can refer to as they build their own systems.

\begin{figure}[t]
\centering
\includegraphics[width=.75\columnwidth]{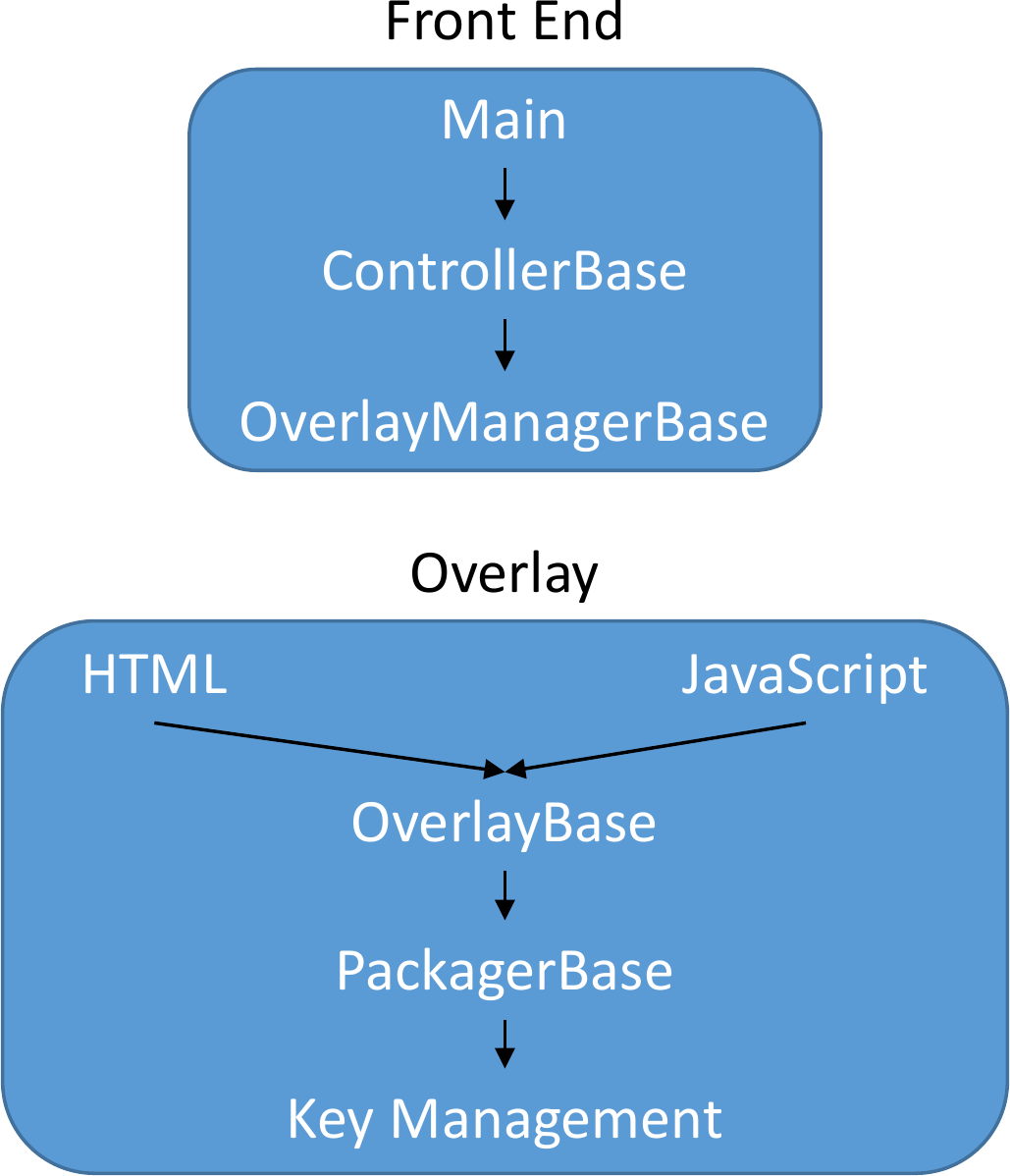}
\caption{\mg's customizable framework.}
\label{fig:framework-objects}
\end{figure}

\mg was designed to minimize the amount of code that must be changed in order for researchers to build new prototypes. The customizable classes enabling this rapid prototyping are shown in Figure~\ref{fig:framework-objects}. \mg includes a default instantiation for each of the base classes (e.g. \texttt{ControllerBase}) seen in the figure.
To change the global functionality of \mg, researchers need to change the aforementioned default implementations.
If researchers desire to implement new functionality (e.g., create a new overlay, support a new application), they can instead subclass these base classes.
All classes, both base classes and default implementations, can be extended, but only allow researchers to override the methods that are unique to their functionality.

\subsection{Frontend}
The main class is responsible for parsing the URL and instantiating the appropriate controller (i.e., classes extending \texttt{ControllerBase}).
Frontend controllers are responsible for the actual operations of the frontend, including detecting when overlays are needed and placing those overlays.
Every overlay is created by and coupled to an overlay manager, which is responsible for handling communication between the overlay and \mg's frontend.
Currently, MessageGuard provides overlay managers for both reading and composing encrypted content.

The simplest way to modify the frontend is to change the elements that it will overlay.
This can be done by changing the CSS selector that is passed to \texttt{ControllerBase}'s constructor.\footnote{
Though unlikely to be necessary, it is also possible to modify the controller to do more complex selection that does not rely on CSS selection.
}
The controller can also be configured to support additional types of overlays (i.e., creating a unified read and compose overlay for instant messaging clients).
In this case, it will also be necessary to create an overlay manager to communicate with the new overlay.

Using these base classes, \mg's default functionality was implemented using less than 200 lines of JavaScript.

\subsection{Overlays}
Overlays are composed of both HTML interfaces and JavaScript code.
Researchers can either modify the existing overlays (read and compose), or create their own overlays.
The steps for creating a new overlay modifying overlays on a per application basis are as follows:
\begin{enumerate}

\item Create a new HTML file for each overlay. This will define the visual appearance of the overlay. 

\item Create a custom read, compose, or entirely new overlay (e.g., file upload) by extending either the \texttt{OverlayBase} class or one the reference overlays (read and compose). These parent classes provide basic functionality (e.g., positioning, communication with the frontend). \item Connect the overlay's HTML interface to its controlling code by referencing this new JavaScript class in the new HTML.

\item Create a new overlay manager to work with the new overlay. You can extend any of the existing overlay managers, or create a new one by extending \texttt{OverlayManagerBase}. 
\item Add any custom communication code to both the new overlay and overlay manager.

\end{enumerate}

\mg's default read overlay required 70 lines of HTML and 150 lines of JavaScript to implement. The default compose overlay needed 190 lines of HTML and 670 lines of JavaScript, most of which was responsible for setting up the HTML5 rich-text interface and allowing users to select a specific key for encryption.

\subsection{Packager}
By overriding PackagerBase, it is possible to create custom message packages, allowing \mg to support a wide range of content-based encryption protocols.
This functionality can be used to allow prototypes developed with \mg to inter-operate with existing cryptographic systems (e.g., using the PGP package syntax in order to be compatible with existing PGP clients).
It could also be used to experiment with advanced cryptographic features, such as key ratcheting~\cite{unger2015sok}.
%In order to deploy a new packager, an existing overlay will need to be modified or a new overlay created which instantiates the custom packager instead of the default packager.

\subsection{Key Management}
Key management is one of the most poorly understand areas of usable, content-based encryption.
While there are many advocates of particular key management schemes (e.g., PGP), there has been little work to actually analyze the empirical usability of these schemes.
One key goal of \mg is to allow existing proposals for key management to be implemented in a real system, and then compared against alternative schemes.
As such, we took special care to ensure that \mg would be compatible with all key management schemes we are currently aware of.
%A schema for the key management component is given in Appendix~\nameref{appx:schema}.

In order to create a new key management scheme, the following two classes must be implemented:

\textbf{KeyScheme.}
The KeyScheme is responsible for handling scheme-specific UI functionality for the key manager (e.g., importing public/private keys, authenticating to a key server). The KeyScheme methods are:

\begin{itemize}

\item \textbf{getUI}
Retrieves a scheme-specific UI that will be included with the KeyUIManager's generic UI. This method is provided with the KeySystem being created/updated and a callback which notifies the KeyUIManager that the KeySystem is ready to be saved.

\item \textbf{handleError}
Modifies an existing KeySystem's UI to allow it to address an error. This method is provided with details about the error, the KeySystem UI to modify, and a callback which notifies the KeyUIManager that the error has been resolved.
Examples of errors include not having a necessary key or expired authentication credentials.

\item \textbf{create} Creates a KeySystem from the scheme-specific UI provided to this method.
% Create a KeySystem from the scheme-specific UI passed to this method.

\item \textbf{update} Updates a KeySystem from the scheme-specific UI provided to this method.
% provided to this method.

\end{itemize}

\textbf{KeySystem.}
A KeySystem is an instantiation of a key management scheme that allows the users to decrypt/sign data for a single identity and encrypt/verify data for any number of identities.\footnote{Key systems which don't support recipients set \texttt{canHaveRecipients} to \texttt{false} and ignore the identity parameters.}
A KeySystem is responsible for performing cryptographic operations with the keys it manages.
Every KeySystem has a fingerprint that uniquely identifies it. The KeySystem methods are:

\begin{itemize}

\item \textbf{serialize/deserialize} Prepares data that is not a part of the KeyAttributes type for storage by the KeyStorage class.

\item \textbf{encrypt}
Encrypts data for the provided identity. Returns the encrypted data along with the fingerprint of the KeySystem that can decrypt it.

\item \textbf{decrypt}
Decrypts the provided data.

\item \textbf{sign}
Signs the provided data.

\item \textbf{verify}
Verifies that the provided signature is valid for the provided data.

\end{itemize}

By default, \mg will allow users to use all available key management schemes, though this can be overridden on a per-prototype basis.

\section{Validation}

We evaluated \mg ability to support usable, content-based encryption research on a wide range of platforms.
Additionally, we measured the performance overhead that \mg creates.
Our results indicate that \mg is compatible with most web applications and has minimal performance overhead.
%Furthermore, we find that the interfaces implemented in \mg are rated as highly usable.
%Finally, our experience demonstrates that \mg guard allows for the rapid creation of new content-based encryption prototypes.

\subsection{Ubiquity}
We tested \mg on major browsers and it worked in all cases:
Desktop -- Chrome, Firefox, Internet Explorer, Opera, and Safari. Android -- Chrome, Firefox, Opera. iOS -- Chrome, Mercury, Safari. 
%It also works on the following mobile browsers: Chrome (iOS and Android), Firefox (Android), Mercury (iOS), Opera (Android, iOS), Safari (iOS).

We tested \mg on the Alexa top 50 web sites.
One of the sites is not a web application (\url{t.co}) and another requires a Chinese phone number in order to use it (\url{weibo.com}).
\mg was able to encrypt data in 47 of the 48 remaining web applications.
The one site that failed (youtube.com) did so because the application removed the comments field when it lost focus, which happens when focus switched to \mg's compose overlay.
We were able to address this problem with a customized front end that required only five lines of code to implement.

These results indicate that researchers should be able to use \mg to research content-based encryption for the web applications of their choice with little difficulty.

\subsection{Performance}
We profiled \mg on several popular web applications and analyzed \mg's impact on load times.
In each case, we started the profiler, reloaded the page, and stopped profiling once the page was loaded.
Our results show that \mg has little impact on page load times and does not degrade the user's experience as they surf the Web: Facebook -- 0.93\%, Gmail -- 2.92\%, Disqus -- 0.54\%, Twitter -- 1.98\%.

\begin{table}[t]
{\centering
\resizebox{\columnwidth}{!}{
\begin{threeparttable}
\begin{tabular}{l|ccc|ccc|}

	Stage & \multicolumn{3}{c|}{Static} & \multicolumn{3}{c|}{Dynamic} \\
	$n$ & 100 & 500 & 1000 & 100 & 500 & 1000 \\

	\midrule
	
	Chrome\tnote{1} 	& 1.14 	& 0.84 	& 0.95 	& 3.17 	& 6.49 	& 11.0 	\\ 
	Firefox\tnote{1}	& 1.06 	& 0.99 	& 0.96 	& 2.26 	& 3.15	& 4.45 	\\
	Safari\tnote{1}	& 0.45 	& 0.63 	& 0.53 	& 3.73 	& 12.8	& 25.5 	\\
	\midrule	
	
	Chrome\tnote{2} 	& 4.27 	& 4.39 	& 4.60	& 12.9 	& 30.2	& 51.1 \\
	Chrome\tnote{3}	& 5.68 	& 5.97 	& 5.94	& 12.4 	& 32.0	& 61.2 \\
	Safari\tnote{3}	& 2.57	& 2.46 	& 1.79	& 15.1 	& 25.2	& 39.5 \\

  \bottomrule
\end{tabular}

{\scriptsize
\begin{tablenotes}
%	\item[1] MacBook Air (OSX 10.10.3, 13-inch, Mid 2013, 1.7GHz Intel Core i7, 8GB RAM). \\Chrome -- 42.0.2311.135, Firefox -- 37.0.2, Safari -- 8.0.5.
	\item[1] MacBook Air (OSX 10.10.3, 1.7GHz Core i7, 8GB RAM). \\Chrome -- 42.0.2311.135, Firefox -- 37.0.2, Safari -- 8.0.5.
	\item[2] OnePlus One (CyanogenMod 12S, AOSP 5.1, 64GB). \\Chrome -- 42.0.2311.47.
	\item[3] iPad Air (iOS 8.3, \nth{1} gen, 64GB). \\Chrome -- 42.0.2311.47, Safari -- 8.0.
\end{tablenotes}
}

\end{threeparttable}
}}

\caption{Average time to overlay an element (ms)}
\label{tab:performance}
\end{table}

Since \mg is intended to work with all websites, we created a synthetic web app that allowed us to test \mg's performance in extreme situations.
This app measures \mg's performance when overlaying static content present at page load (Stage 1) and when overlaying dynamic content that is added to the page after load (Stage 2).
The application takes as input $n$, the number elements that will be overlayed in each stage.
Half of these elements will require read overlays and half will require compose overlays.

%The synthetic app has two phases:
%The first phase tests \mg's performance when overlaying static content and the second phase test \mg's performance when overlaying dynamically added content.
%The application takes as input the number of overlays ($n$) that should be added in each phase.
%It then creates $\nicefrac{n}{4}$ \texttt{TextAreas}, $\nicefrac{n}{4}$ \texttt{ContentEditable Divs}, and $\nicefrac{2n}{4}$ encrypted payloads.
%\mg is then initialized and we record the time taken to overlay this initial set of items.
%Next, an additional $\nicefrac{n}{4}$ \texttt{TextAreas}, $\nicefrac{n}{4}$ \texttt{ContentEditable Divs}, and $\nicefrac{2n}{4}$ encrypted payloads are iteratively added to the page and we record the time taken to overlay this new content.
%%In total, $2n$ overlays are added to the page.

Using this synthetic web application, we tested \mg with six browsers and three values of $n$.
We averaged measurements over ten runs and report our findings in Table~\ref{tab:performance}.
Performance for overlaying static content does not significantly vary based on the number of overlays created.
In contrast, performance for overlaying dynamic content for most browsers seems to grow polynomial in the number of overlays added.
Still, performance in the Firefox desktop browser demonstrates that this is not an inherent limitation of \mg.
Finally, we note that even in extreme cases (dynamic - $n=1000$) overlaying occurs quickly (max 61 ms).

\mg's low performance overhead indicates it is suitable for building responsive prototypes, suitable for testing by users.
Moreover, if performance problems arise, researchers can be reasonably sure that the problems are in their changes to \mg.

\section{Case Studies}
While developing \mg, we developed two content-based encryption prototypes using \mg: a secure Facebook Chat prototype and a secure email prototype.
In this section we describe these prototypes, the effort taken to build them, and IRB-approved usability studies of these prototypes.
These case studies demonstrate that \mg enables building high quality research prototypes that are well-liked by users.

\subsection{Private Facebook Chat}
Using an early version of \mg, we created Private Facebook Chat (PFC), a system that adds content-based encryption to chat on Facebook~\cite{robison2012private}.
PFC leverages identity-based encryption (IBE) in order to transparently manage encryption keys, removing the need for users to establish shared secrets or obtain public keys in before sending chat messages.

\begin{figure}[t]
\centering
\includegraphics{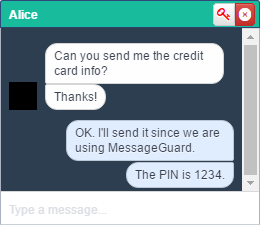}
\caption{A unified overlay for Facebook Chat.}
\label{fig:facebook}
\end{figure}

PFC uses a custom controller that detects the Facebook Chat interface.
Instead of having separate read and compose overlays, PFC uses a single unified overlay that handles displaying and composing encrypted messages (see Figure~\ref{fig:facebook}).
Using a single overlay was advantageous because it was more space efficient and better mimicked a typical instant messaging interface.\footnote{In general, we recommend this approach for instant messaging clients.}

We conducted an IRB-approved user study of PFC involving 17 participants.
Almost all users were able to use PFC to encrypt their chat sessions, except two extremely novice users that were unable to complete any tasks.
Additionally, users indicated that they were generally satisfied with PFC and that they would be interested in using it in practice.

The Facebook Chat prototype took a single student working part-time one month to complete (about 80 hours).
Most of this time was spent designing the interface for the unified overlay and modifying \mg's IBE key management scheme to support authentication through Facebook Connect.
In total, PFC's implementation required 50 lines of HTML and 350 lines of JavaScript.

\subsection{Private Webmail (Pwm)}

The design of \mg was guided by our development of a series of secure email prototypes for Gmail \cite{ruoti2013confused,ruoti2015why,ruoti2016private} that we used to study usable, secure email for the masses.
These prototypes all go by the name Private Webmail (Pwm).
Pwm uses identity-based encryption (IBE) to allow users to send encrypted email to recipients who have not yet installed Pwm, something that is not possible with PGP or S/MIME.
In addition, Pwm focuses on helping first time users understand how to decrypt their first secure email and how to use Pwm to encrypt messages.

\begin{figure}[t]
\centering
\includegraphics[width=\columnwidth]{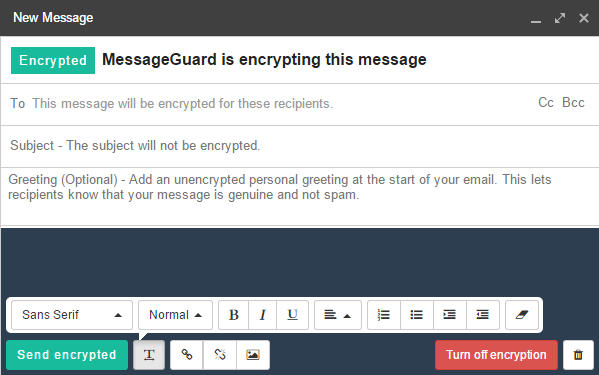}
\caption{Customized \mg front end for email.}
\label{fig:email-integrated}
\end{figure}

In order to maximize Pwm's integration with Gmail we created a customized front end that annotates Gmail's interface.
Our specialized front annotates Gmail's interface to help users understand better how their email is protected (see Figure~\ref{fig:email-integrated}).
We also added the ability for users to include a plaintext greeting with their encrypted emails, helping recipients feel safe when receiving encrypted email.
Additionally, we added context-sensitive, inline tutorials that instructed users on how to use Pwm and what protections it affords their email.
%With the generic front end, Gmail's \textbf{To} field has a button to enable encryption because it is a \texttt{TextArea}, even though this doesn't make sense in the context of email.

We have evaluated Pwm across five different IRB-approved usability studies, including a total of 186 participants.
Each of these studies utilized a standard usability metric, the System Usability Scale (SUS)~\cite{bangor2008empirical,bangor2009determining}, 
which generates a single score between 0 and 100 that is an indication of a system's usability.
SUS consists of ten discriminating questions that users answer after completing tasks using \mg.

The first three user studies all involved the same version of Pwm, with the first study (25 participants) having participants install Pwm using a bookmarklet, and the second (32 participants) and third study (28 participants) having participants install Pwm using an extension~\cite{ruoti2013confused}.
In each of the three studies, recipients were told to wait for an email containing information needed to continue the study, though they were not informed that this email would be encrypted.
They were then sent the encrypted email, and were required to decrypt this email, before continuing to send and reply to several more encrypted emails.
Overall, Pwm averaged a SUS score of 73.8, putting it in the \nth{70} percentile of systems tested with SUS --- first study (75.7), second study (70.7), third study (70.7).

Qualitative feedback from these three studies revealed that we had made the encryption and key management details too transparent and users were unsure whether or not to trust the system.
This caused us to modify our design to make some encryption details more apparent in the interface~\cite{ruoti2016private}.
Using this modified version of Pwm, we conducted our fourth usability study.
It involved 51 participants that experimented with the bookmarklet version of \mg and incorporated changes based on the results of prior studies.
The results of this test was a SUS score of 80.0, falling in the \nth{88} percentile of systems tested with SUS.
More importantly, our modifications to Pwm (which were also incorporated into \mg's reference implementation) succeeded in addressing user concerns.
Nearly all participants (92\%) believed that their friends and family could easily start using it. 

Our fifth study examined whether Pwm could be adopted in a grassroots fashion~\cite{ruoti2016we}.
To test this, we brought in pairs of novice users, and instructed the first participant to send an encrypted email to the second participant.
The first participant was only given a link to Pwm, while the second user was not informed that they would be using encrypted email.
In this study, Pwm received a SUS score of 72.3, falling in the \nth{63} percentile of systems tested with SUS.\footnote{In this study we also examined other systems, all of which received uncharacteristically low scores, suggesting that Pwm's true SUS score is much closer to the 80 found in our fourth study.}
Furthermore, all users successfully used Pwm to begin transmitting encrypted emails between themselves, and most users praised Pwm's tutorials and intuitive interfaces.

As Pwm was developed in lock-step with \mg, it is difficult to know exactly how much time was spent specifically on Pwm specific functionality, though we estimate around 150 hours.
Most of this time was spent reverse engineering Gmail's interface, which is difficult because is dynamically generated and minimizes both HTML and JavaScript.
We also spent significant time creating and refining Pwm's tutorials.
In total, Pwm includes approximately 1300 news lines of JavaScript.

\subsection{Lessons Learned}

Both case studies demonstrate that it is possible to rapidly build content-based encryption prototypes using \mg.
While these prototypes still required some time to implement (80 and 150 hours), this pales in comparison to the nearly 2000 hours that have already gone into the development of \mg.
Our work with Pwm especially helped us see the value of comparing competing systems; lab usability studies have been
particularly helpful in helping us to differentiate between the usability of several different secure email systems.
Finally, these case studies demonstrate that \mg can be used to build prototypes that are rated by users as being highly usable.
This is especially important in the case of Pwm, as Pwm's interfaces are highly similar to the interfaces used in \mg's default functionality.

\section{Conclusion}

We described \mg, a platform for usable security research focused on content-based encryption.
\mg is designed to encourage collaboration between the security and usability research communities
to solve longstanding usability problems in this space. It simplifies development of prototypes and
comparison of competing designs, while also providing a platform for sharing and replicating
research results. \mg retrofits existing applications, providing broad utility to developers,
and includes secure cryptographic operations so that usability is tested on functioning systems.
We validated the performance and deployability of \mg and shared several case studies demonstrating
its utility for the field.
Our hope is that \mg will help security and usability researchers to
cooperate in solving numerous pressing problems and speed the adoption
of usable, content-based encryption.

\section{Acknowledgment}
This work was partially funded by Sandia National Laboratories.
Sandia National Laboratories is a multi-program laboratory managed and operated by Sandia Corporation, a wholly owned subsidiary of Lockheed Martin Corporation, for the U.S. Department of Energy's National Nuclear Security Administration under contract DE-AC04-94AL85000.
We would like to thank Jay McCarthy, Paul van Oorschot, and Dan Olsen for providing feedback on this paper.

\balance

\bibliographystyle{IEEEtran}
\bibliography{paper}

%\clearpage
%\setcounter{figure}{0}
%\setcounter{table}{0}
\appendix

\subsection{Implementation Strategies}
\label{appx:strategies}

To determine the appropriate implementation strategy for \mg, we examined approaches for enhancing web applications with content-based encryption.
In this section, analyze these approaches with respect to their deployability, security, and usability.
We group the implementation strategies into two categories: integrated strategies and non-integrated strategies.
We also propose two new hybrid strategies that address limitations in the existing strategies and are better suited to meet our system goals.

\subsubsection{Integrated Strategies}
%\begin{figure}[t]
%\centering
%\includegraphics[width=1.0\columnwidth]{figures/overlays.png}
%
%\vspace{.5\baselineskip}
%Portions of the web application (shown on the left) have been overlayed with secure interfaces (shown on the right).
%
%\caption{Overlaying a web application}
%\label{fig:overlays-orig}
%\end{figure}

Integrated strategies attempt to add content-based encryption directly to the interfaces of existing websites.
In a process known as \textbf{overlaying}~\cite{van2009encrypted}, the browser can replace portions of the website's interface with secure interfaces known as \textbf{overlays}.% (Figure~\ref{fig:overlays-orig}).
These overlays are displayed as if they were a part of the web application, but their contents are inaccessible to the web application.
Often, there are different overlays for different functions, such as composing textual content, uploading files, and decrypting content.

\textbf{iFrame.}
The oldest integrated strategy is to create overlays using HTML iFrames~\cite{van2009encrypted,ruoti2013confused}.
Since iFrames are a part of the browser's DOM, it is trivial to integrate them with websites.
Further, iFrames are protected by the browser's same-origin policy.
There are three methods for overlaying websites with iFrames:

\begin{enumerate}

\item \textbf{Browser extension.} Extensions are desktop only, except in the case of Firefox on Android.
%The source of the extension must be trusted, similar to an app store.

\item \textbf{Bookmarklet.}
Bookmarklets are user scripts that are stored as browser bookmarks.
When a user clicks a bookmarklet, the associated script is executed on the current page.
Bookmarklets are supported on all platforms, both desktop and mobile, and do not require users to install any software.

When using bookmarklets, the iFrame's \texttt{src} attribute will need to reference a trusted remote domain.
To maximize security, this domain should only host the contents needed by the bookmarklet's user script.
This limited functionality makes it easier to lock down the domain's servers in order to prevent an intrusion by the adversary.

Bookmarklets are affected by the browser's Content Security Policy (CSP), which can be used to limit the source of frames and scripts.
Still, this limitation has been marked as a bug in Chromium\footnote{\url{https://code.google.com/p/chromium/issues/detail?id=233903}} and Firefox,\footnote{\url{https://bugzilla.mozilla.org/show_bug.cgi?id=866522}} and it is possible that in the future this limitation will be removed.
In practice, only a small number of websites use CSP. %TODO: Get a citation here.

\item \textbf{Proxy application.}
A proxy application can be used to augment the bookmarklet approach.
First, it can modify the CSP settings of a website to allow iFrame's to reference the system's trusted domain.
Second, a proxy can automatically inject the bookmarklet script into a webpage, obviating the need for the user to click the bookmarklet.
Third, a proxy can allow the iFrame strategy to work with non-browser applications that display HTML interface elements retrieved from the web.
A proxy application can work on any platform, except for Windows on mobile devices.
Still, there are two significant drawbacks to this approach: (1) for non-rooted phones, the proxy must be implemented using the phone's VPN service and (2) it must proxy HTTPS connections in order to modify their contents.

\end{enumerate}

In prior work, we used iFrames to develop Pwm, a secure email client for the masses~\cite{ruoti2013confused}.
Pwm tightly integrated with Gmail and was designed to maximize usability.
Pwm focused on helping first time recipients of an encrypted email understand how to install Pwm and decrypt their message.
Pwm was rated as highly usable by participants in several usability studies.
Since our work on Pwm, additional systems in both research~\cite{atwater2015leading} and industry (e.g., Virtru, Mailvelope) have also used the iFrame strategy to secure email.

\textbf{Shadow DOM.}
Shadow DOM is a new feature proposed in the HTML5 specification, and currently has a partial implementation in Blink-based browsers (e.g., Chrome).
The Shadow DOM allows for the creation of a ``shadow'' \texttt{DocumentFragment} (i.e., \texttt{ShadowRoot}) that will be rendered in place of another fragment (i.e., host) on the website.
% in the actual DOM.
Elements contained in the \texttt{ShadowRoot} are invisible to the rest of the DOM and must be accessed through their parent \texttt{ShadowRoot}.
For example, \texttt{document.getElementById} cannot find children of a \texttt{ShadowRoot}.
%but can be retrieved by calling \texttt{getElementById} on their parent  \texttt{ShadowRoot}.

Overlays can be implemented using a \texttt{ShadowRoot}.
However, while \texttt{ShadowRoot} objects and their contents should only be accessible to the entity that created the \texttt{ShadowRoot} object, in practice this is not the case:

\begin{enumerate}
\item The Shadow DOM allows for an element to have multiple \texttt{ShadowRoots}, with the newest \texttt{ShadowRoot} having access to older \texttt{ShadowRoot} objects through their \texttt{olderShadowRoot} property.

\item There are two CSS selectors that ``pierce'' the Shadow DOM. First the ``$>>>$'' selector can be used to grab any element that matches the selectors to the right of ``$>>>$'', regardless of whether that element is inside a \texttt{ShadowRoot}. Similarly, the ``::shadow'' pseudo-selector can be used to select the \texttt{ShadowRoot} attached to any element.

\item The \texttt{Element.prototype.createShadowRoot} method\hfill\break can be replaced by the web application with a version that saves references to the created \texttt{ShadowRoots}, allowing the web application to access the contents of these \texttt{ShadowRoots}.
% in the web application, which can use the references to access the contents of the \texttt{ShadowRoots}.

\end{enumerate}

In order to implement overlays using a \texttt{ShadowRoot}, it is necessary to modify the JavaScript environment to address these access methods~\cite{he2014shadowcrypt}.
Because of this limitation, ShadowDOM cannot be implemented using a bookmarklet, which cannot guarantee that the JavaScript environment is modified before the website has the ability to store pointers to these access methods.
Instead, a Shadow DOM strategy must be implemented using either an extension or a proxy application.

He et al. used the Shadow DOM to implement ShadowCrypt, a Chrome extension that attempted to add secure communication to arbitrary websites~\cite{he2014shadowcrypt}.
In our experience, usability issues keep ShadowCrypt from working with most websites.
Moreover, flaws in ShadowCrypt's implementation allow the website to retrieve the user's sensitive data.\footnote{A detailed description of this problem is given later in the appendix.}
This demonstrates the danger of relying on non-standard security models created by developers.

\textbf{Greasemonkey.}
Greasemonkey is a Firefox plugin that allows for websites to be modified with custom interface elements defined using XUL.
XUL interface elements can be used for overlays, and the contents of these interfaces are protected by the browser's sandboxing of websites~\cite{chromiumSandbox}.

Fahl et al. used Greasemonkey to create a Firefox extension that added content-based encryption to Dropbox, Facebook, and email~\cite{fahl2012confidentiality}.
Their plugin used Confidentiality-as-a-Service (CaaS) to automate key management.
Usability studies demonstrate that their system was highly usable.

\textbf{Accessibility.}
\label{sec:accessibility}
Most operating systems have an accessibility layer that is used by accessibility tools to modify apps to make them usable by individuals with disabilities.
Lau et al. proposed using the accessibility layer to implement secure overlays~\cite{lau2014mimesis}.
In this strategy, the content of secure overlays is protected by the browser's sandbox.
While our work is concerned with encrypting content within the browser, the accessibility layer is interesting in that it has the potential to secure non-browser applications.
While iFrames implemented using a proxy application can also support non-browser applications, the accessibility layer has the potential to be more universal and reliable.

Lau et al. used this strategy to develop Mimesis Aegis (M-Aegis), a system for enhancing Android apps with content-based encryption.
%privacy-preserving system for Android.
%M-Aegis is the first system that attempts to provide ubiquitous and integrated encryption outside of the browser.
Their system secured several communication apps (e.g., Gmail) and also provided a method for support to be added for additional apps.\footnote{This method is not generic and requires custom logic for each supported application.}
A usability study of the Android Gmail app secured with M-Aegis showed that most participants did not report any noticeable difference between the original Gmail app and the Gmail app with M-Aegis enabled.

\subsubsection{Non-integrated Strategies}
In contrast to integrated strategies, non-integrated strategies use separate applications to provide encryption, relying on the user to copy-and-paste encrypted content in and out of the original application.
Because these applications are not integrated with websites, they are free to make UI design choices that maximize usability, without any concern that these choices will clash with a website's UI.
%Still, research has shown that users strongly prefer encryption interfaces that integrate with the websites they are protecting~\cite{ruoti2013confused,atwater2015leading}.

\textbf{Standalone website.}
A standalone website can be used to handle encryption and decryption.
While it is not integrated with websites, it is integrated with the browser.
Since a browser is used to run the website, no installation is required.
Security is provided by the browser's same-origin policy.
This approach is not common, but there have been a few standalone websites that provide PGP encryption.\footnote{For example, \url{https://www.igolder.com/pgp/encryption/} and \url{http://www.hanewin.net/encrypt/PGcrypt.htm}.}

\textbf{Standalone Application.}
A standalone application is the traditional strategy for implementing content-based encryption.
Security is provided by the operating system's app separation policies.

The earliest example of a standalone content-based encryption application is Pretty Good Privacy (PGP). 
PGP was created in 1991 by Phil Zimmerman and is used to protect a wide range of data.
Since then, many similar programs have been created.

\subsubsection{Comparison}

\begin{table*}
%\fontsize{2.5mm}{0.8em}\selectfont,
\centering
\begin{threeparttable}
\begin{tabular}{l|cccccc|ccc|ccc|ccc|l|}

	&
		\multicolumn{6}{c|}{Browsers} &
		\multicolumn{3}{c|}{Desktop OS} &
		\multicolumn{3}{c|}{Mobile OS} &
		\multicolumn{3}{c|}{Usability} &
	\\

	Strategy &
		\rot{Chrome} & \rot{Firefox} & \rot{IE} & \rot{Opera} & \rot{Safari} & \rot{Others} &
		\rot{Linux} & \rot{Mac OS} & \rot{Windows} &
		\rot{Android} & \rot{iOS} & \rot{Windows} &
		\rot{Integration} & \rot{\shortstack[l]{Non-standard\\Interfaces}} & \rot{No install} & 
		Security Model
	\\
	\toprule

	iFrame &
		\full & \full & \full & \full & \full & \full &
		\full & \full & \full &
		\half & \half & \half\tnote{5} &
		\full & \half & \half\tnote{5} & 
		Same-origin
	\\
	
	Shadow DOM  &
		\half\tnote{2} & \none & \none & \half\tnote{2} & \none & \none &
		\full & \full & \full &
		\half & \half & \none &
		\full & \half & \none & 
		\textcolor{red}{None}
	\\
	
	Greasemonkey\tnote{1} &
		\none & \full & \none & \none & \none & \none &
		\full & \full & \full &
		\half & \none & \none &
		\full & \half & \none & 
		Browser Sandbox
	\\	
	
	\midrule	
	Accessibility &
		\full & \full & \full & \full & \full & \half\tnote{3} &
		\none\tnote{4} & \half & \half &
		\half & \none & \none &
		\full & \half & \none & 
		Browser Sandbox
	\\
	
	\midrule
	Standalone website &
		\full & \full & \full & \full & \full & \full &
		\full & \full & \full &
		\full & \full & \full &
		\none\tnote{6} & \full & \full & 
		Same-origin
	\\
	
	Standalone App &
		\full & \full & \full & \full & \full & \full &
		\half & \half & \half &
		\half & \half & \half &
		\none & \full & \none & 
		Browser Sandbox
	\\
		
  \bottomrule
\end{tabular}

\full~Full support, \half~Partial support, \none~No support

\begin{tablenotes}
\begin{multicols}{2}
	\item[1] Scheduled for deprecation.
	\item[2] ShadowDOM is not fully supported in Chrome.
	\item[3] Depends on whether the browser is accessible.
	\item[4] In development -- AT-SPI~\cite{atspi}.
	\item[5] Potentially limited by CSP.
	\item[6] Integrated with the browser, but not the website.
\end{multicols}
\end{tablenotes}
\end{threeparttable}
\caption{Comparison of implementation strategies}
\label{tab:systems}
\end{table*}

We have analyzed and compared each of the above strategies based on their deployability, usability, and security. The results are summarized in Table~\ref{tab:systems}.

\textbf{Deployability.}
There are three key areas of deployability: which browsers are supported, which desktop operating systems are supported, and which mobile devices are supported.
Both the Greasemonkey and Shadow DOM strategies are limited to a single type of browser, Firefox and Blink-based, respectively.
On mobile, Greasemonkey only works with Firefox on Android and the Shadow DOM requires the use of a proxy application, which as mentioned earlier is less than ideal (marked as ``Partial support'' in Table~\ref{tab:systems}).
Greasemonkey uses XUL, which is slated for deprecation.
On the other hand, Shadow DOM is part of the HTML5 specification and there is a good chance that sometime in the future it will become standard in more browsers.

The deployment challenge faced by both the accessibility and standalone app strategy is that every platform, both desktop and mobile, requires its own implementation.
For standalone apps, this burden could be reduced through the use of a cross-platform framework (e.g., Java, Mono), though these frameworks commonly lead to systems with a poor look-and-feel.
Unfortunately, there is not a cross-platform accessibility layer and each platform does require a unique implementation.
These limitations have been marked in Table~\ref{tab:systems} using the ``Partial support'' symbol.

Both iFrames and standalone websites are built on commonly deployed approaches, and work on all platforms and browsers.
iFrames on mobile can be implemented using either bookmarklets or an application proxy. In the case of bookmarklets, iFrames are limited by websites' CSP policies, and in the case of a proxy application on mobile devices, there is the degradation of user experience (marked as ``Partial support'' in Table~\ref{tab:systems}).

\textbf{Usability.}
The standalone strategies suffer from a lack of integration, which is disliked by users~\cite{ruoti2013confused,atwater2015leading}.
On the other hand, because the standalone strategies are not tied to the website's interface, the standalone strategies support non-standard interfaces (e.g., input field drawn using an HTML canvas).
While the integrated strategies could contain logic to handle website-specific interfaces, this approach is a significant engineering effort (marked as ``Partial support'' in Table~\ref{tab:systems}).

The standalone website strategy does not require installation.
When the iFrame strategy is implemented using bookmarklets, there is also no installation.
%, but bookmarklets are limited by CSP.

\textbf{Security.}
\label{sec:security}
The Greasemonkey, accessibility, and standalone app strategies all rely on the browser's sandbox for security~\cite{chromiumSandbox}.
The sandbox prevents websites from being able to access any resources outside of the websites DOM, including the browser chrome (Greasemonkey) and other applications (accessibility, standalone app).
While the browser's sandbox has been compromised in the past, such attacks are quickly patched~\cite{pwn2own}.

The iFrame and standalone website strategies rely on the browser's same-origin policy for security~\cite{barth2011web,securityHandbook}.
As part of this policy, browsers ensure that code executing on an arbitrary website is unable to access or modify content hosted on a different domain (i.e., iFrame, standalone website).
This is an important model that provides security for much of the web~\cite{stefan2014protecting}.

Similar to compromises of the browser sandbox, from time to time the same-origin policy is partially broken.
For example, in 2013 Paul Stone described an attack that used CSS and measurements of render times to leak the contents of an iFrame~\cite{stone2013pixel}.
After disclosure of this flaw, the browser vendors worked with Stone and quickly addressed the issue.
This is the advantage of relying on a security model actively supported by browsers: problems are quickly fixed.

\begin{figure}[t]

\begin{lstlisting}
var elements = document.querySelectorAll('[contentEditable]');
for (var i = 0, len = elements.length; i < len; i++) {
	var newShadowRoot = elements[i].createShadowRoot();
	newShadowRoot.innerHTML = '<shadow></shadow>';
	if (newShadowRoot.olderShadowRoot)
		console.log(newShadowRoot.olderShadowRoot.querySelector('textarea').value);
}
\end{lstlisting}

\textbf{Attack 1.} Steals content from ShadowCrypt elements by creating a new \texttt{ShadowRoot} and using it to access ShadowCrypt's \texttt{ShadowRoot}. ShadowCrypt stores content in a \texttt{TextArea} and so we extract it from there.
%Modifies the DOM. 
This attack is possible because ShadowCrypt deletes the \texttt{Document.createShadowRoot} function instead of the \texttt{Document.prototype.createShadowRoot} function.

\vspace{\baselineskip}

\begin{lstlisting}
var textArea = document.querySelectorAll('*::shadow textarea');
for (var i = 0, len = textArea.length; i < len; i++) {
  console.log(textArea[i].value);
}
\end{lstlisting}

\textbf{Attack 2.} Steals content from ShadowCrypt elements by using the \texttt{::shadow} pseudo-selector.
%Does not modify the DOM.
This attack is possible because ShadowCrypt does not filter the ``::shadow'' pseudo-selector. Due to frequent changes in the Shadow DOM specification, it is likely this selector did not exist when ShadowCrypt was implemented.

\caption{Attacks on ShadowCrypt}
\label{fig:attacks}

\end{figure}

\begin{table*}
%\fontsize{2.5mm}{0.8em}\selectfont,
\centering
\begin{tabular}{l|cccccc|ccc|ccc|ccc|l|}

	&
		\multicolumn{6}{c|}{Browsers} &
		\multicolumn{3}{c|}{Desktop OS} &
		\multicolumn{3}{c|}{Mobile OS} &
		\multicolumn{3}{c|}{Usability} &
	\\

	Strategy &
		\rot{Chrome} & \rot{Firefox} & \rot{IE} & \rot{Opera} & \rot{Safari} & \rot{Others} &
		\rot{Linux} & \rot{Mac OS} & \rot{Windows} &
		\rot{Android} & \rot{iOS} & \rot{Windows} &
		\rot{Integration} & \rot{\shortstack[l]{Non-standard\\Interfaces}} & \rot{No install} & 
		Security Model
	\\
	\toprule

	\shortstack[l]{iFrame +\\Standalone website} &
		\full & \full & \full & \full & \full & \full &
		\full & \full & \full &
		\full & \full & \full &
		\full & \full & \full & 
		Same-origin
	\\

	\midrule
	\shortstack[l]{Accessibility +\\Standalone App} &
		\full & \full & \full & \full & \full & \half &
		\none & \half & \half &
		\half & \none & \none &
		\full & \full & \full & 
		App Separation
	\\
		
  \bottomrule
\end{tabular}

\vspace{\baselineskip}
\caption{Comparison of hybrid strategies}
\label{tab:hybrid}
\end{table*}

Shadow DOM is the one approach that does not rely on a standard security model, but instead requires developers to modify the JavaScript environment to prevent websites from accessing the contents of a \texttt{ShadowRoot}.
The dangers of this approach can be seen through a security analysis we conducted of ShadowCrypt.
We downloaded the latest version of ShadowCrypt\footnote{Version 0.3.3, released February 4, 2015.} and found two attacks that allowed web applications to read plaintext data stored in the secure overlays.
The JavaScript to run these attacks and explanation of why they work are given in Figure~\ref{fig:attacks}

These attacks demonstrate two problems with relying on developers to add security to the Shadow DOM.
First, it is difficult for developers to correctly identify all attack vectors and correctly close them (Figure~\ref{fig:attacks} -- Attack 1).
Second, as browsers are updated, it is possible that new attack vectors are added, and developers must be constantly vigilant (Figure~\ref{fig:attacks} -- Attack 2).

\textbf{JavaScript-based Cryptography.}
The iFrame, Shadow DOM, Greasemonkey, and standalone website strategies all rely upon cryptographic primitives implemented in JavaScript.
There have been concerns that in certain cases JavaScript-based cryptography is untrustworthy~\cite{javascriptHarmful}.
These arguments reduce to two different concerns.
First, if cryptography is being used because TLS is not trusted by the website, then the website cannot guarantee that the JavaScript is delivered to the user's browser unmodified.
Second, if cryptography is being used to encrypt the data so that it is opaque to the website, then you cannot trust the website to send you JavaScript that will encrypt this data.

Both of these are valid concerns, but are orthogonal to the strategies that use JavaScript-based cryptography.
First, the strategies \textit{do} trust TLS to correctly deliver the content-based encryption software.
Second, the strategies \textit{do not} trust the website, and that is precisely why the encryption software is separate from the website.
Finally, except in the case of iFrames implemented using Bookmarklets, the cryptographic JavaScript code is only downloaded once, at installation (iFrame, Shadow DOM, Greasemonkey) or on first run (standalone website).

\subsubsection{Hybrid Strategies}

The comparison of implementation strategies shows that the iFrames and accessibility strategies best match our goals of retrofitting the web with content-based encryption.
Still, both of these approaches struggle with non-standard interfaces.
To address this, we suggest that each of these strategies be modified to fall back to a standalone strategy when users encounter a non-standard interface that the system was unable to overlay.
%For example, when a non-standard interface is encountered by a user, they could click a button on the system that would launch a standalone content-based encryption interface.
We pair strategies that have the same security model: i.e., iFrame + standalone website (same-origin) and Accessibility + standalone app (browser sandbox).
The capabilities of these new hybrid approaches are summarized in Table~\ref{tab:hybrid}.

Using the hybrid strategies, a bookmarklet implementation of iFrames can protect websites that don't employ CSP, and for those websites that do use CSP, the implementation can fall back to the standalone website.
This standalone website could be displayed in either a new window or tab.
Alternatively, when a user encounters a non-standard interface that is not overlayed, they could click a button to launch the standalone website.
This allows iFrames to fully support mobile platforms without using a proxy application.
Additionally, this means that the no-install implementation of iFrames (i.e., bookmarklets) is available for all browsers and platforms.

\subsubsection{Best Strategy}
Based on our goal of retrofitting the web with content-based encryption, the hybrid strategy \textbf{iFrames + standalone website} is the best strategy.
It works with all current browsers and platforms, both mobile and desktop.
Furthermore, in most situations it provides tight integration with websites, but can fall back to a standalone website when non-standard interfaces are encountered.
Moreover, using bookmarklets, this strategy can be run on systems where the user does not have install permissions.

%The accessibility + standalone app strategy has significant deployability limitations and is not ideal for our goal.
%Still, the ability for this strategy to secure applications other than the browser is very interesting.
%This strategy could become the optimal strategy if the following two events occurred:
%First, a single platform-neutral accessibility framework needs to be implemented across all operating systems, reducing the development burden for this strategy (e.g., AT-SPI~\cite{atspi}).
%Second, iOS and mobile Windows need to allow for the installation of accessibility tool apps.

\subsection{Implementation}
\label{appx:implementation}
In this section, we describe in greater depth the technical implementation of \mg.

\subsubsection{Front End}
When initialized, the front end immediately scans the page using the documents \texttt{querySelectorAll} and a \texttt{TreeWalker}.
After this initial scan, changes to the page are tracked using a single \texttt{MutationObserver} and only elements that have been modified are scanned.
This process allows \mg to have a minimal effect on page load times and application execution.

Where possible, the front end uses the Shadow DOM to position overlays (i.e., styling, not security).
When the Shadow DOM is unavailable, as is the case in most browsers, we instead set the overlay's style to match the position and size of the overlayed element, and then set the overlayed element's \texttt{display} style to \texttt{none}.
This alternative approach has some potential to interfere with the underlying application (e.g., \texttt{:nth-child()}) and is not as desirable as using the Shadow DOM.

\subsubsection{Cryptography}
Where ever possible stark2009symmetric used Node.js's Crypto library.
For functionality not provided by this library, we used the Stanford Javascript Crypto Library~\cite{stark2009symmetric} (SJCL).
The only cryptographic primitives that we had to develop ourselves were for the implementation of Boneh-Boyen IBE.
This was necessary as there no publicly available implementations IBE that were suitable for inclusion in \mg.

\subsubsection{Browser Extension}
We developed the \mg browser extension using Kango, a cross-browser extension framework.
Kango packages \mg and makes it available as a browser extension for Chrome, Firefox, Opera, and Safari.
Within a browser extension, the front end is injected into a web application before the web application is loaded.
The overlays, packager, and key management components operate within the extension's trusted origin, protecting them from the web application.

In Safari, browser extensions are subject to the Content Security Policy (CSP), which prevents \mg from functioning on sites that set a \texttt{frame-src} CSP attribute.
Until March 10 of this year, the CSP specification explicitly disallowed applying CSP policies to extensions and bookmarklets, but the language has since been weakened to allow browsers to choose whether CSP protections apply to extensions and bookmarklets~\cite{spec:csp}.
Since Safari's broken functionality existed before March 10 and because all other browsers exempt extensions from CSP, we believe it is likely Safari will eventually exempt extensions as well.
%Regardless, to address cases where CSP causes difficulty, \mg's options page includes a compose overlay and read overlay where users can copy and paste text in order to have it encrypted or decrypted.

\subsubsection{Bookmarklet}
Bookmarklets are user scripts that are stored as browser bookmarks.
When a user clicks a bookmarklet, the associated script is executed on the current page.
We have implemented \mg so that it can be executed from a bookmarklet.
This is helpful since mobile browsers do not currently support extensions but do support bookmarklets.

The browser extension and bookmarklets share the same codebase and are nearly identical.
The only significant difference is that \mg's components are hosted from a standard web origin (e.g., https://messageguard.com) instead of the local extension origin.
If \mg's origin was compromised, it would be possible for an attacker to inject malicious scripts into user's browsers.
To help mitigate this, we recommend that \mg's bookmarklet be hosted on an origin with no other responsibilities, allowing this origin to be significantly locked down.

Bookmarklets are not currently exempted from CSP protections, but this has already been marked as a bug in Chromium\footnote{\url{https://code.google.com/p/chromium/issues/detail?id=233903}} and Firefox.\footnote{\url{https://bugzilla.mozilla.org/show_bug.cgi?id=866522}}
%As with the extension, the \mg options page can be used to encrypt/decrypt sensitive data if problems occur.

%\section{Key Management Schema}
%\label{appx:schema}
%\begin{figure*}[t]
%\centering
%\includegraphics[width=1.0\textwidth]{figures/key-framework.pdf}
%
%\vspace{.5\baselineskip}
%A new key management scheme can be added by implementing the KeyScheme and KeySystem classes.
%\caption{Schema of the Key Management Framework.}
%\label{fig:key-framework}
%\end{figure*}
%
%See Figure~\ref{fig:key-framework}.

\end{document}